\documentclass[preprint,preprintnumbers,nofootinbib]{revtex4-1}

\usepackage[dvips]{color}
\usepackage{graphicx}
\usepackage{amsmath}
\usepackage{amssymb}
\usepackage{amsfonts}

\newcommand{\re}{{\rm Re}}

\newcommand{\diag}{{\rm Diag}}
\newcommand{\ev}{{\rm eV}}

\newcommand{\gev}{{\rm GeV}}

\newcommand{\bmx}{\left(\begin{array}}
\newcommand{\emx}{\end{array}\right)}

\begin{document}

\vspace{2cm}
\preprint{MISC-2013-03}

\title{Is $\theta_{13}^{\rm PMNS}$ correlated with $\theta_{23}^{\rm PMNS}$ or not?}
\author{
Takeshi Araki\footnote{araki@cc.kyoto-su.ac.jp}
}
\affiliation{
Maskawa Institute, Kyoto Sangyo University,\\ Motoyama, Kamigamo, Kita-Ku, Kyoto 603-8555, Japan
}

\begin{abstract}
By postulating the relation $\theta_{23}^{} \simeq 45^\circ + \eta\theta_{13}^{}$, we seek preferable correction terms to tri-bi-maximal mixing and discuss their origins.
Global analyses of the neutrino oscillation parameters favor $\eta=\pm 1/\sqrt{2}$; this corresponds to the relation found by Edy, Frampton, and Matsuzaki some years ago in the context of a $T^\prime$ flavor symmetry.
In contrast, the results of the $\nu_\mu^{}$ disappearance mode reported by the T2K and Super-Kamiokande collaborations seem to prefer $\eta=0$, which gives an almost maximal $\theta_{23}^{}$.
We derive a general condition for ensuring $\theta_{23}^{} \simeq 45^\circ + \eta\theta_{13}^{}$ and find that the condition is complicated by the neutrino masses and CP violating phases.
We investigate the condition under simplified environments and arrive at several correction terms to the mass matrices.
It is found that the obtained correction terms can arise from flavor symmetries or one-loop radiative corrections.
\end{abstract}

\maketitle

\section{Introduction}
Now that $\theta_{13}^{}$ of the PMNS mixing matrix has been measured very precisely by reactor \cite{rct} and long-baseline \cite{acl} neutrino oscillation experiments, it may be said that we have succeeded in acquiring a clear picture of the neutrino mixing pattern.
It is the Daya-Bay experiment that holds the record of precise determination of $\theta_{13}^{}$: the vanishing $\theta_{13}^{}$ is now excluded at the level of $7.7\sigma$ standard deviations with an unexpectedly large central value, $\theta_{13}^{\rm DB} \simeq 8.7^\circ$ \cite{DAYA}.
Such a large $\theta_{13}^{}$ would offer a great opportunity for us to explore the neutrino mass ordering, octant of $\theta_{23}^{}$, and leptonic CP violation.

On the theoretical side, the discovery of the large $\theta_{13}^{}$ disappointed many people, since this could signal the end of a paradigm of tri-bi-maximal (TBM) mixing \cite{TBM}, which predicts $\theta_{13}^{}=0^\circ$.
Nevertheless, TBM mixing may still be useful as a leading order one in the presence of small corrections.
In fact, various ways to complement the TBM mixing by perturbing the neutrino sector \cite{prtb-nu,max23}, the charged lepton sector \cite{prtb-ch}, or both \cite{prtb-both} have recently been proposed.
In this work, we take the same stance while paying special attention to a certain correlation between $\theta_{13}^{}$ and $\theta_{23}^{}$.

Within the framework of TBM mixing plus small corrections, deviations of $\theta_{13}^{}$ and $\theta_{23}^{}$ from their TBM values are given by the same parameters, as we shall show in Eqs. (\ref{eq:t23}) and (\ref{eq:s13}); thus it is likely that they are somehow correlated with each other.
Indeed, that often happens in some flavor models \cite{corr}.
Given these facts, it is very intriguing to recall the relation $\theta_{13}^{}=\sqrt{2}|45^\circ - \theta_{23}^{}|$ found by Edy, Frampton and Matsuzaki in the context of a $T^\prime$ flavor symmetry \cite{efm} (see also Ref. \cite{efm-org} for earlier works).
We will hereafter refer to this relation as the EFM relation.
Interestingly, the EFM relation is now in excellent agreement with the global fits of the neutrino oscillation parameters, as shown in Table \ref{tab:efm}.
\begin{table}[h]
\begin{center}
\begin{tabular}{|c||c|c|c|}\hline
 Data & $\theta_{23}^{\rm best}$ & $\theta_{13}^{\rm best}$ & $\theta_{13}^{\rm EFM}$ \\ \hline
 Ref. \cite{fogli} & $38.4^\circ(38.7^\circ)$ & $8.9^\circ(9.0^\circ)$ & $9.3^\circ(8.9^\circ)$ \\ \hline
 Ref. \cite{valle} & $51.5^\circ(50.8^\circ)$ & $9.0^\circ(9.1^\circ)$ & $9.2^\circ(8.2^\circ)$ \\ \hline
\end{tabular}
\end{center}
\caption{Comparisons of the EFM predictions with the global fits of the neutrino oscillation parameters in the case of normal (inverted) mass ordering: $\theta_{13}^{\rm best}$ and $\theta_{23}^{\rm best}$ are the best fit values from Refs. \cite{fogli} and \cite{valle}, while $\theta_{13}^{\rm EFM}$ is a prediction of the EFM relation when employing $\theta_{23}^{\rm best}$.}\label{tab:efm}
\end{table}
Such a simple relation is expected to help us obtain an insight into building a successful flavor model.

In contrast, the results of the $\nu_\mu^{}$ disappearance mode reported by the T2K \cite{t2k-moriond} and Super-Kamiokande \cite{atm-nu12} collaborations still seem to favor the maximal $\theta_{23}^{}$, indicating that only $\theta_{13}^{}$ departs from the TBM value independently of $\theta_{23}^{}$.
If this is the case, it will be important to figure out the origin of stability of $\theta_{23}^{}$ while inducing an appreciable deviation for $\theta_{13}^{}$.

In view of these thoughts, in this work, we seek preferable correction terms to the charged lepton and neutrino mass matrices with a guide of
\begin{eqnarray}
\theta_{23}^{} \simeq 45^\circ + \eta \theta_{13}^{},
\label{eq:rela}
\end{eqnarray}
where $\eta=\pm 1/\sqrt{2}$ ends up as the EFM relation, while $\eta=0$ corresponds to the case of an almost maximal $\theta_{23}^{}$.
Note that we use $\simeq$ in Eq. (\ref{eq:rela}), because some approximations are used in our discussions.
After showing our definitions of mass and mixing matrices in Sec. II, we derive a general condition for realizing Eq. (\ref{eq:rela}) in Sec. III.
The obtained condition is somewhat complicated by the neutrino masses and the CP violating phases.
Thus, in Sec. IV, we investigate the condition for a specific neutrino mass spectrum and/or CP violating phases and show several examples of the correction terms.
In Sec. V, we develop two possible ways to realize the correction terms obtained in Sec. IV by means of flavor symmetries or radiative corrections.
We summarize our discussions in Sec. VI.

\section{Definitions}
We begin with the SM Lagrangian augmented by an effective Majorana mass term for the left-handed neutrinos:
\begin{eqnarray}
{\cal L}={\cal L}_{\rm SM}^{} 
- \nu^\dag_L M_\nu^{} \nu^{*}_L + h.c.~,
\end{eqnarray}
and consider circumstances under which $\theta_{13}^{}$ and $\theta_{23}^{}$ are required to be $0^\circ$ and $45^\circ$, respectively, by an underlying flavor physics at leading order.
Up to Sec. IV, $\theta_{12}^{}$ remains arbitrary in order to keep our discussions as general as possible.
The leading-order PMNS matrix is defined as $U^0=(U^0_\ell)^\dag U^0_\nu = V^0 P^0$, where $U^0_\ell$ and $U^0_\nu$ stand for the leading-order diagonalizing matrices of the charged leptons and the neutrinos, respectively, $P^0=\diag(e^{i\alpha/2},e^{i\beta/2},1)$ and
\begin{eqnarray}
V^0=
\bmx{ccc}
c_{12}^0 & s_{12}^0 & 0 \\
-s_{12}^0/\sqrt{2} & c_{12}^0/\sqrt{2} & -1/\sqrt{2} \\
-s_{12}^0/\sqrt{2} & c_{12}^0/\sqrt{2} &  1/\sqrt{2}
\emx .\label{eq:v0}
\end{eqnarray}
$\alpha$ and $\beta$ are Majorana CP violating phases, and $s_{12}^0$ (or $c_{12}^0$) denotes $\sin\theta_{12}^0$ (or $\cos\theta_{12}^0$).
We divide mass matrices into leading and small correction terms:
\begin{eqnarray}
M_\ell^{}=M_\ell^0 + \delta M_\ell^{},~~~
M_\nu^{}=M_\nu^0 + \delta M_\nu^{} ,\nonumber
\end{eqnarray}
where $M_\ell^{}$ is a mass matrix of the charged leptons.
$M_\ell^0$ and $M_\nu^0$ are diagonalized by $U_\ell^0$ and $U_\nu^0$, respectively, whereas $\delta M_\ell$ and $\delta M_\nu$ give rise to small corrections to $U^0_{}$.
Approximating the diagonalization of $M_\ell^{}$ as
\begin{eqnarray}
(U_\ell^0+\delta U_\ell^{})^\dag M_\ell^{} M_\ell^\dag (U_\ell^0+\delta U_\ell^{})
\simeq \diag(m_e^2,m_\mu^2,m_\tau^2), \nonumber
\end{eqnarray}
we move on to the diagonal basis of $M_\ell^{}$ and redefine $M_\nu^{}$ as follows:
\begin{eqnarray}
M_\nu \rightarrow \bar{M}_\nu 
&=& (U_\ell^0+\delta U_\ell^{})^\dag(M_\nu^0 + \delta M_\nu^{})(U_\ell^0+\delta U_\ell^{})^*
\nonumber \\
&\simeq& (U_\ell^0)^\dag M_\nu^0 (U_\ell^0)^* + \delta U_\ell^\dag M_\nu^0 (U_\ell^0)^* + (U_\ell^0)^\dag M_\nu^0 \delta U_\ell^* + (U_\ell^0)^\dag \delta M_\nu^{} (U_\ell^0)^*
\nonumber \\
&\equiv& \bar{M}_\nu^0 + \delta \bar{M}_\nu^{} , \label{eq:Mn}
\end{eqnarray}
where $\bar{M}_\nu^0 = (U_\ell^0)^\dag M_\nu^0 (U_\ell^0)^*$ and we have dropped several terms in the second line.
We stress that $\delta \bar{M}_\nu$ includes corrections stemming from not only the neutrino sector but also the charged lepton sector.

$\bar{M}_{\nu}^{0}$ is expressed in terms of the leading-order mixing angles and masses as $\bar{M}_\nu^0 = V^0_{} {\cal D}_\nu^0 (V^0_{})^T_{}$ where ${\cal D}_\nu^0 = \diag(m_1^0e^{i\alpha} ,m_2^0e^{i\beta},m_3^0)$.
Meanwhile, $\delta \bar{M}_\nu$ is described by three complex parameters; we parametrize it as follows:
\begin{eqnarray}
\delta \bar{M}_\nu^{}
=V^0_{}
\bmx{ccc}
0 & X & Y \\
X & 0 & Z \\
Y & Z & 0
\emx (V^0_{})^T
,\label{eq:xyz}
\end{eqnarray}
where the diagonal entries are omitted since they can be absorbed into ${\cal D}_\nu^0$.

\section{Relating $\theta_{13}^{}$ with $\theta_{23}^{}$}
We consider $\bar{M}_\nu^{} \bar{M}_\nu^\dag$\footnote{In the first paper of Ref. \cite{max23}, a perturbation method is also adopted, but for $\bar{M}_\nu^{}$ in view of $V^*_{ij}\simeq V^{}_{ij}$. As we shall explain in Sec. IV-B, we will arrive at the same conclusion.}:
\begin{eqnarray}
\bar{M}_\nu^{} \bar{M}_\nu^\dag 
\simeq \bar{M}_\nu^0 (\bar{M}_\nu^0)^\dag + \bar{M}_\nu^0 (\delta\bar{M}_\nu^{})^\dag + \delta\bar{M}_\nu^{}(\bar{M}_\nu^0)^\dag ,\nonumber
\end{eqnarray}
and regard the second and third terms as small perturbations to $\bar{M}_\nu^0 (\bar{M}_\nu^0)^\dag$.
Then, the perturbed mixing angles are found to be
\begin{eqnarray}
&&\tan\theta_{12}
\simeq
t_{12}^0\left[ 
1+\frac{1}{s_{12}^0 c_{12}^0}\re \frac{m_1^0 e^{i\alpha}X^* + m_2^0 e^{-i\beta}X}{(m_2^0)^2 - (m_1^0)^2}
\right], \label{eq:t12}\\
&&\tan\theta_{23}
\simeq 1+ 2\re 
\left[
\frac{m_1^0 e^{i\alpha}Y^* + m_3^0 Y}{(m_3^0)^2 - (m_1^0)^2}s_{12}^0 - \frac{m_2^0 e^{i\beta}Z^* + m_3^0 Z}{(m_3^0)^2 - (m_2^0)^2}c_{12}^0
\right], \label{eq:t23}\\
&&\sin\theta_{13}
=\left| 
\frac{m_1^0 e^{i\alpha}Y^* + m_3^0 Y}{(m_3^0)^2 - (m_1^0)^2}c_{12}^0 + \frac{m_2^0 e^{i\beta}Z^* + m_3^0 Z}{(m_3^0)^2 - (m_2^0)^2}s_{12}^0 
\right|, \label{eq:s13}
\end{eqnarray}
whereas there are no corrections to the eigenvalues up to the first order expansion.
It can be seen that $X$ in Eq. (\ref{eq:xyz}) is mainly responsible for deviations of $\theta_{12}$ while those of $\theta_{23}$ and $\theta_{13}$ are controlled by $Y$ and $Z$.

As for $\theta_{13}$ and $\theta_{23}$, we are particularly interested in the relation $\theta_{23} \simeq 45^\circ + \eta\theta_{13}$.
In view of $\cos\theta_{13}\simeq 1$, this relation may be translated into 
\begin{eqnarray}
\sin\theta_{23}=\sin[45^\circ + \eta\theta_{13}]
\simeq \frac{1}{\sqrt{2}}(1 + \eta\sin\theta_{13}),\nonumber
\end{eqnarray}
leading us to
\begin{eqnarray}
&&\eta\left| 
\frac{m_1^0 e^{i\alpha} Y^* + m_3^0 Y}{(m_3^0)^2 - (m_1^0)^2}c_{12}^0 + \frac{m_2^0 e^{i\beta} Z^* + m_3^0 Z}{(m_3^0)^2 - (m_2^0)^2}s_{12}^0 
\right| \nonumber \\
&&\hspace{2cm}= 
\re\left[  
\frac{m_1^0 e^{i\alpha} Y^* + m_3^0 Y}{(m_3^0)^2 - (m_1^0)^2}s_{12}^0 - \frac{m_2^0 e^{i\beta} Z^* + m_3^0 Z}{(m_3^0)^2 - (m_2^0)^2}c_{12}^0
\right].\label{eq:13-23}
\end{eqnarray}
This is the condition for obtaining $\theta_{23} \simeq 45^\circ + \eta\theta_{13}$ and is one of our main results in this study.
Unfortunately, the condition contains undetermined observables, i.e. the individual neutrino mass and the Majorana phases, and looks complex.
Hence, we will investigate the condition in some special cases in the next section.

\section{Deviations from TBM mixing}
\begin{figure}[t]
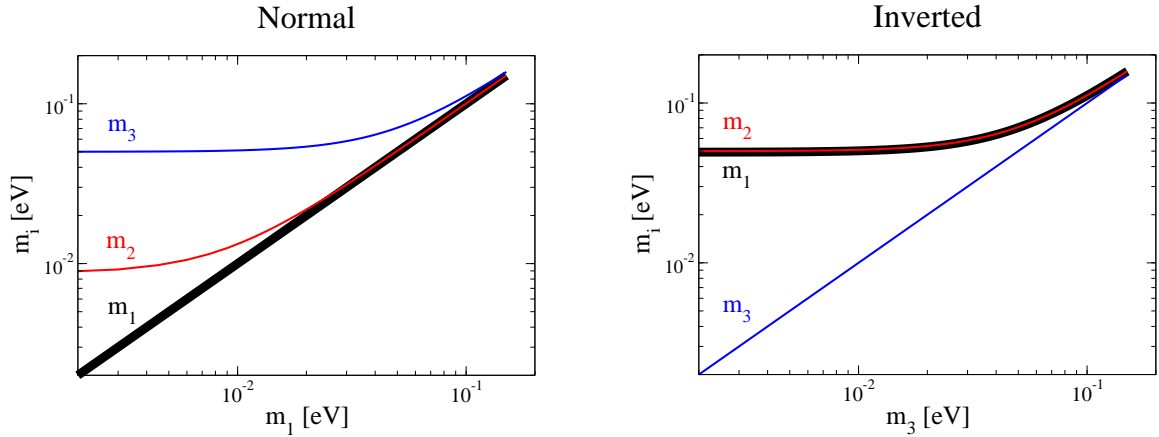

\begin{center}
\includegraphics[width=7.0cm]{running-NH.eps}
\hspace{1.0cm}
\includegraphics[width=7.0cm]{running-IH.eps}
\end{center}
\caption{\footnotesize 
The mass spectrum of neutrinos as a function of the lightest neutrino mass for the normal (left panel) and inverted (right panel) ordering cases.
Best fit values are used for $\Delta m_{12}^2$ and $\Delta m_{23}^2$.
The black, red, and blue curves correspond to $m_1^{}$, $m_2^{}$, and $m_3^{}$, respectively.
} \label{fig:mass-running}
\end{figure}

From now on, we concentrate on the case of $s_{12}^0=1/\sqrt{3}$; namely, $V^0$ in Eq. (\ref{eq:v0}) takes the form of the TBM mixing pattern.
Moreover, we will postulate $(m_3^0)^2 - (m_1^0)^2 = (m_3^0)^2 - (m_2^0)^2$.
As mentioned just below Eq. (\ref{eq:s13}), the eigenvalues are only moderately corrected, so it should be a good approximation to identify $m_i^0$ with the physical neutrino masses $m_i^{}$.
Therefore, in the light of $\Delta m_{12}^2 \ll \Delta m_{23}^2$, demanding $(m_3^0)^2 - (m_1^0)^2 = (m_3^0)^2 - (m_2^0)^2$ is expected to be reasonable.
Even with this simplification, however, Eq. (\ref{eq:13-23}) remains complicated.
Thus, in what follows, we will make several assumptions for the neutrino masses and the CP phases.
For reference purposes, we depict the mass spectrum of neutrinos as a function of the lightest one in Fig. \ref{fig:mass-running}.
In all of our numerical calculations, the following best fit values and/or $1\sigma$ errors from Ref. \cite{fogli} are used:
\begin{eqnarray}
&&\Delta m_{23}^2=
\left\{\begin{array}{l}
m_3^2 - m_2^2 = (2.43^{+0.06}_{-0.10})\times 10^{-3} \hspace{1cm}{\rm for\ Normal \ ordering} \\
m_2^2 - m_3^2 = (2.42^{+0.07}_{-0.11})\times 10^{-3} \hspace{1cm}{\rm for\ Inverted\  ordering}
\end{array}\right. , \nonumber \\
&&\Delta m_{12}^2 = (7.54^{+0.26}_{-0.22})\times 10^{-5}_{},~~~~
\sin^2\theta_{12}^{}= 0.307^{+0.018}_{-0.016}~.
\end{eqnarray}
Parameter spaces adopted in the calculations are summarized in Table \ref{tab:para}.
\begin{table}[t]
\begin{center}
\begin{tabular}{|c|c||c|c|c|c|c|}\hline
 & & $m_1^{}~[\ev]$ & $|Y|~[\ev]$ & $\arg Y$ & $\alpha$ & $\beta$ \\ \hline
Fig. 2 & $Z=0$ & $0 \sim 0.1$ & $0 \sim 0.003$ & $0$ & $0,~\pi$ & $=\alpha$ \\ \cline{2-7}
       & $Z=2\sqrt{2}Y$ & $0 \sim 0.1$ & $0 \sim 0.001$ & $0$ & $0,~\pi$ & $=\alpha$ \\ \hline
Fig. 3 & $Z=0$ & $0, ~0.03, ~0.1$ & $0 \sim 0.01$ & $0 \sim 2\pi$ & $0$ & $0$ \\ \cline{2-7}
       & $Z=2\sqrt{2}Y$ & $0, ~0.03, ~0.1$ & $0 \sim 0.003$ & $0 \sim 2\pi$ & $0$ & $0$ \\ \hline
Fig. 4 & $Z=0$ & $0, ~0.03, ~0.1$ & $0\sim 0.015$ & $0 \sim 2\pi$ & $0$ & $0$ \\ \cline{2-7}
      & $Z=2\sqrt{2}Y$ & $0, ~0.03, ~0.1$ & $0\sim 0.003$ & $0 \sim 2\pi$ & $0$ & $0$ \\ \hline
Fig. 5 & $Z=0$ & $0$ & $0\sim 0.011$ & $0$ & $0 \sim 2\pi$ & $0 \sim 2\pi$ \\ \cline{2-7}
      & $Z=2\sqrt{2}Y$ & $0$ & $0\sim 0.003$ & $0$ & $0 \sim 2\pi$ & $0 \sim 2\pi$ \\ \hline
Fig. 6 &  & $0,~0.05\sim 0.1$ & $0\sim 0.007$ & $0\sim 2\pi$ & $0 \sim 2\pi$ & $=\alpha$ \\ \hline
Fig. 7 &  & $0\sim 0.1$ & $0\sim 0.013$ & $\pi/2$ & $0,~\pi$ & $0,~\pi$ \\ \hline
\end{tabular}
\end{center}
\caption{A summary of the parameter spaces used in the numerical calculations.
$m_2^{}$, $m_3^{}$ and $X$ are suitably tuned to be consistent with the $1\sigma$ constraints of $\Delta m_{12}^2$, $\Delta m_{23}^2$, and $\theta_{12}^{}$.}
\label{tab:para}
\end{table}

\subsection{$\eta=\pm 1/\sqrt{2}$}

\begin{itemize}
\item Case A-I: $Y=Y^*$, $Z=Z^*$, and $\alpha=\beta=0$ or $\pi$.\\
Assuming that $Y$ and $Z$ are real parameters and that $\alpha = \beta$ is equal to $0$ or $\pi$, Eq. (\ref{eq:13-23}) is solved to be
\begin{eqnarray}
Z \simeq \frac{\kappa - \eta\sqrt{2}}{\eta + \sqrt{2}\kappa}~ Y
\label{eq:efm-I}
\end{eqnarray}
with $[Y-\sqrt{2}Z]/\eta > 0$ ($<0$) for the case of normal (inverted\footnote{Note that a minus sign appears in the right-hand side of Eq. (\ref{eq:13-23}) in the case of inverted mass ordering.}) mass ordering.
In deriving Eq. (\ref{eq:efm-I}), we have assumed $\pm m_1^{} + m_3^{} = \pm m_2^{} + m_3^{}$.
The validity of this approximation is subject to the neutrino mass spectrum.
When $m_1^{} \simeq m_2^{}$, it is obviously applicable.
Furthermore, the approximation appears valid even in mass regions where $m_1^{}$ is much smaller than $m_2^{}$ in the case of normal ordering, because $m_3^{} \gg m_{1,2}^{}$.
However, as we shall see below, the difference between $m_1^{}$ and $m_2^{}$ gives rise to slight errors, and it makes the EFM relation somewhat hazy.
$\kappa=\pm 1$ stems from a sign ambiguity originating in the left-hand side of Eq. (\ref{eq:13-23}) and provides us with two possible solutions:
\begin{eqnarray}
Z=0~~&{\rm with}&~~Y>0~(Y<0)
\label{eq:dMn-efm-I-Z0}\\
Z=2\sqrt{2}Y~~&{\rm with}&~~Y>0~(Y<0)
\label{eq:dMn-efm-I-ZY}
\end{eqnarray}
for $\eta=1/\sqrt{2}$ ($\eta=-1/\sqrt{2}$) in the case of normal mass ordering.
The sign of $Y$ is flipped for the inverted ordering case.
Note that in the $Z=0$ case, the dependence of Eq. (\ref{eq:13-23}) on the Majorana phase $\beta$ disappears. Therefore, $\beta$ can actually take any value in this case.

\begin{figure}[t]
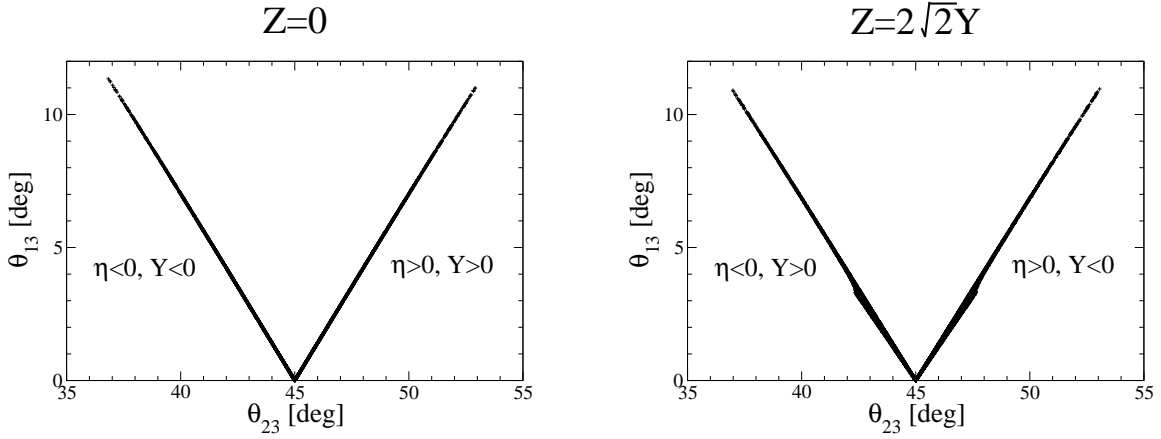

\begin{center}
\includegraphics[width=7.0cm]{atm-rct-efm-I-Z0-NH.eps}
\hspace{1.0cm}
\includegraphics[width=7.0cm]{atm-rct-efm-I-ZY-NH.eps}
\end{center}
\caption{\footnotesize 
$\theta_{13}^{}$ as a function of $\theta_{23}^{}$ for $Z=0$ (left panel) and $Z=2\sqrt{2}Y$ (right panel) in the case of normal mass ordering for Case A-I.
$\Delta m_{12}^2$, $\Delta m_{23}^2$, and $\theta_{12}^{}$ are restricted to be within the 1$\sigma$ bounds.
} \label{fig:atm-rct-efm-I}
\end{figure}
In Fig. \ref{fig:atm-rct-efm-I}, we numerically diagonalize the neutrino mass matrix by imposing Eqs. (\ref{eq:dMn-efm-I-Z0}) and (\ref{eq:dMn-efm-I-ZY}) and compute $\theta_{13}^{}$ as a function of $\theta_{23}^{}$ for the case of normal mass ordering.
In the case of $Z=2\sqrt{2}Y$, the EFM relation is slightly hazy due to the errors from the approximation of $\pm m_1^{} + m_3^{} = \pm m_2^{} + m_3^{}$ when $m_1^{}$ is small.
In contrast, the $Z=0$ case does not suffer from the errors since it does not rely on the approximation.
Almost the same figures are obtained for the inverted ordering case, but such haziness does not show up since $m_1^{}$ is always close to $m_2^{}$.
\item Case A-II: $\alpha=\beta=0$ and $m_1^{}\simeq m_2^{}\simeq m_3^{}$.\\
Eqs. (\ref{eq:efm-I}), (\ref{eq:dMn-efm-I-Z0}), and (\ref{eq:dMn-efm-I-ZY}) are actually valid even for complex $Y$ and $Z$ in the case of the quasi-degenerate neutrino mass spectrum when $\alpha=\beta=0$.
This is because the imaginary parts inside $|\cdots|$ and $\re [\cdots]$ on the left- and right-hand sides of Eq. (\ref{eq:13-23}) are canceled out, and only the real parts are constrained to satisfy
\begin{eqnarray}
\re Z = \frac{\kappa - \eta\sqrt{2}}{\eta + \sqrt{2}\kappa}~ \re Y
\label{eq:efm-II}
\end{eqnarray}
with $[\re Y - \sqrt{2}\re Z]/\eta >0$ ($<0$) for the case of normal (inverted) mass ordering.
Obviously, Eq. (\ref{eq:efm-II}) is automatically satisfied once Eq. (\ref{eq:efm-I}) is imposed for complex $Y$ and $Z$.

\begin{figure}[t]
\begin{center}
\includegraphics[width=7.3cm]{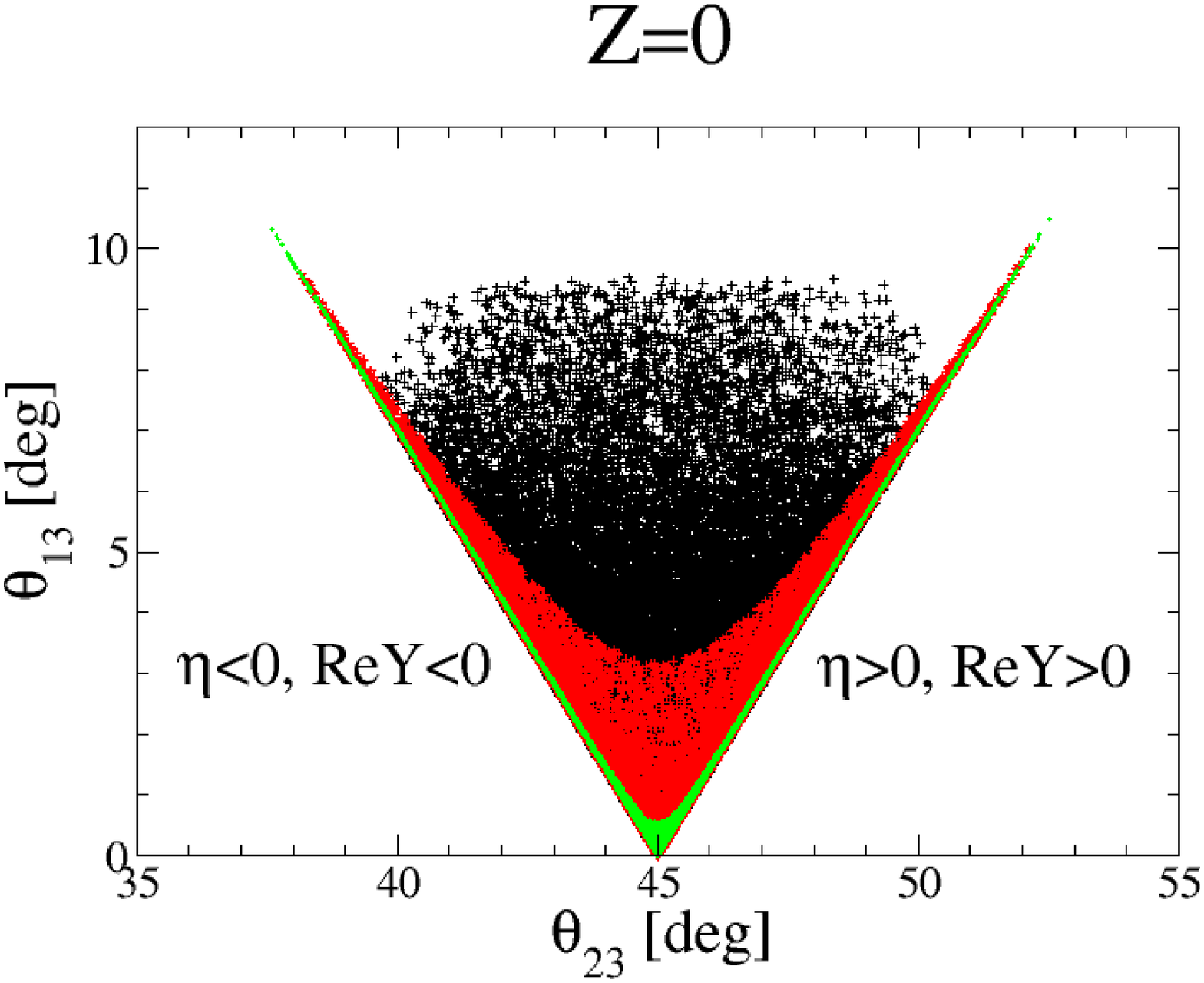}
\hspace{1.0cm}
\includegraphics[width=7.3cm]{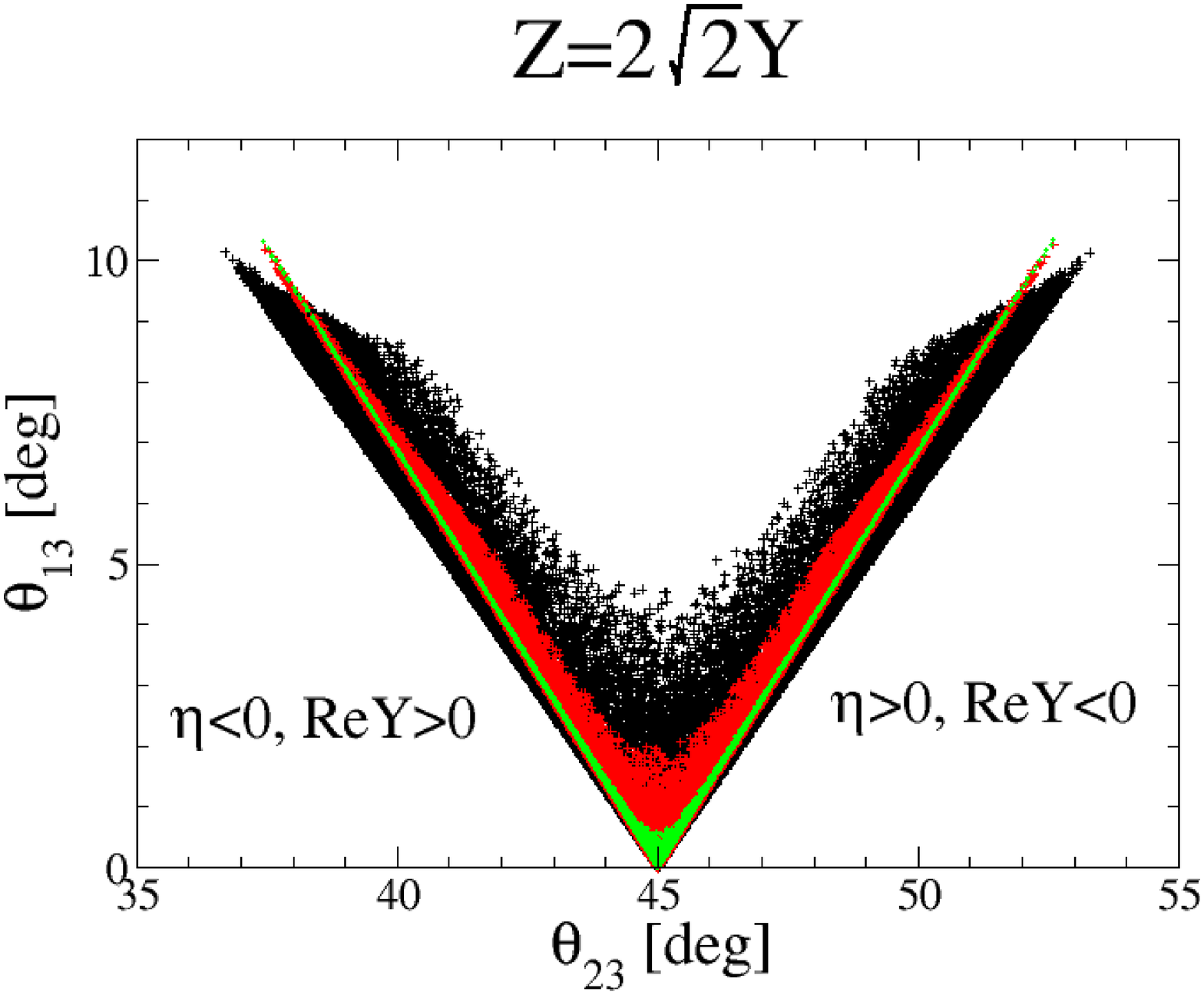} 
\end{center}
\caption{\footnotesize 
$\theta_{13}^{}$ as a function of $\theta_{23}^{}$ for $Z=0$ (left panel) and $Z=2\sqrt{2}Y$ (right panel) in the case of normal mass ordering for Case A-II.
$\Delta m_{12}^2$, $\Delta m_{23}^2$ and $\theta_{12}^{}$ are restricted to be within the 1$\sigma$ bounds.
The black, red, and green dots correspond to $m_{1}^{}=0$, $0.03$, and $0.1~\ev$, respectively.
} \label{fig:atm-rct-efm-II-NH}
%
\begin{center}
\includegraphics[width=7.3cm]{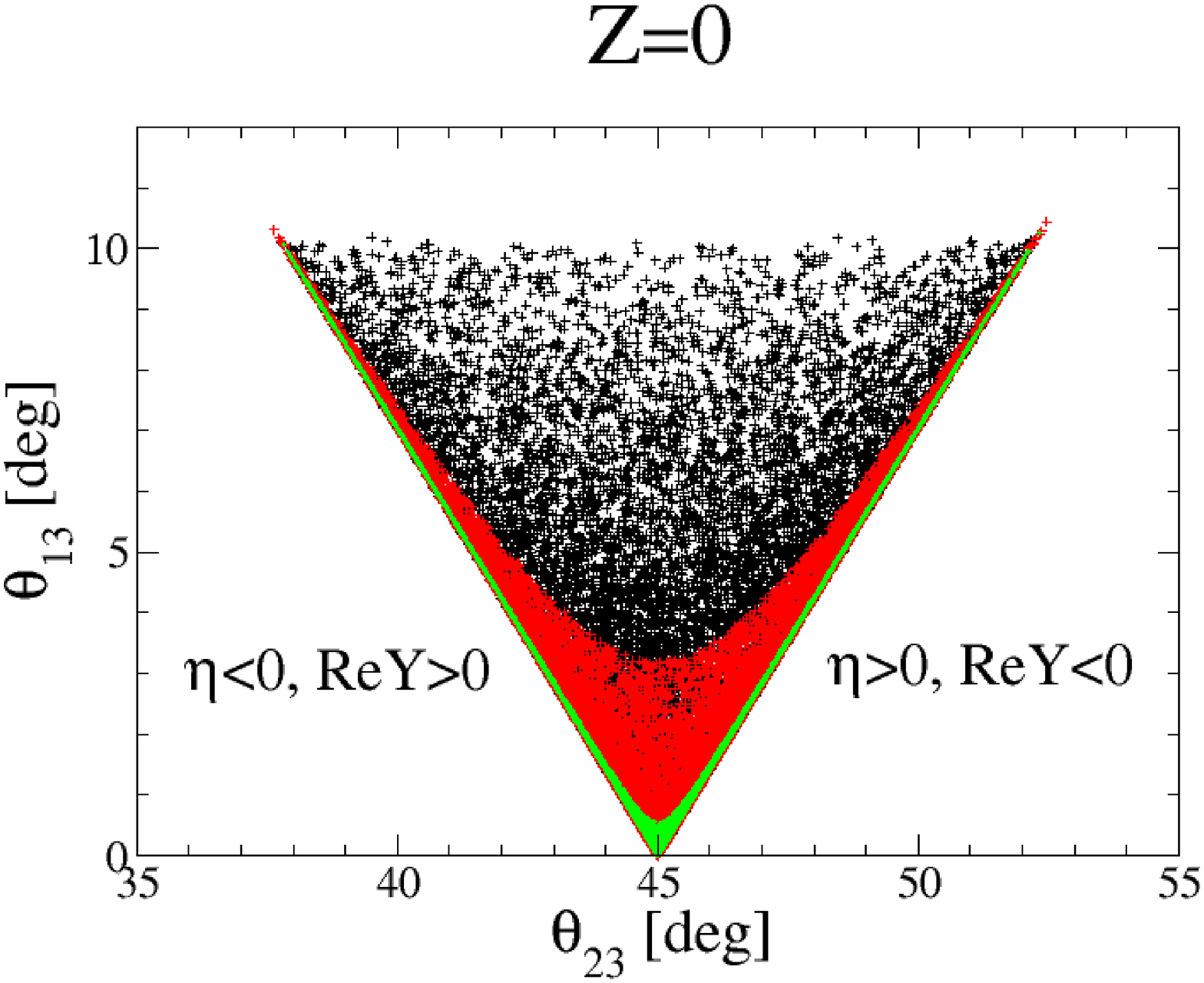}
\hspace{1.0cm}
\includegraphics[width=7.3cm]{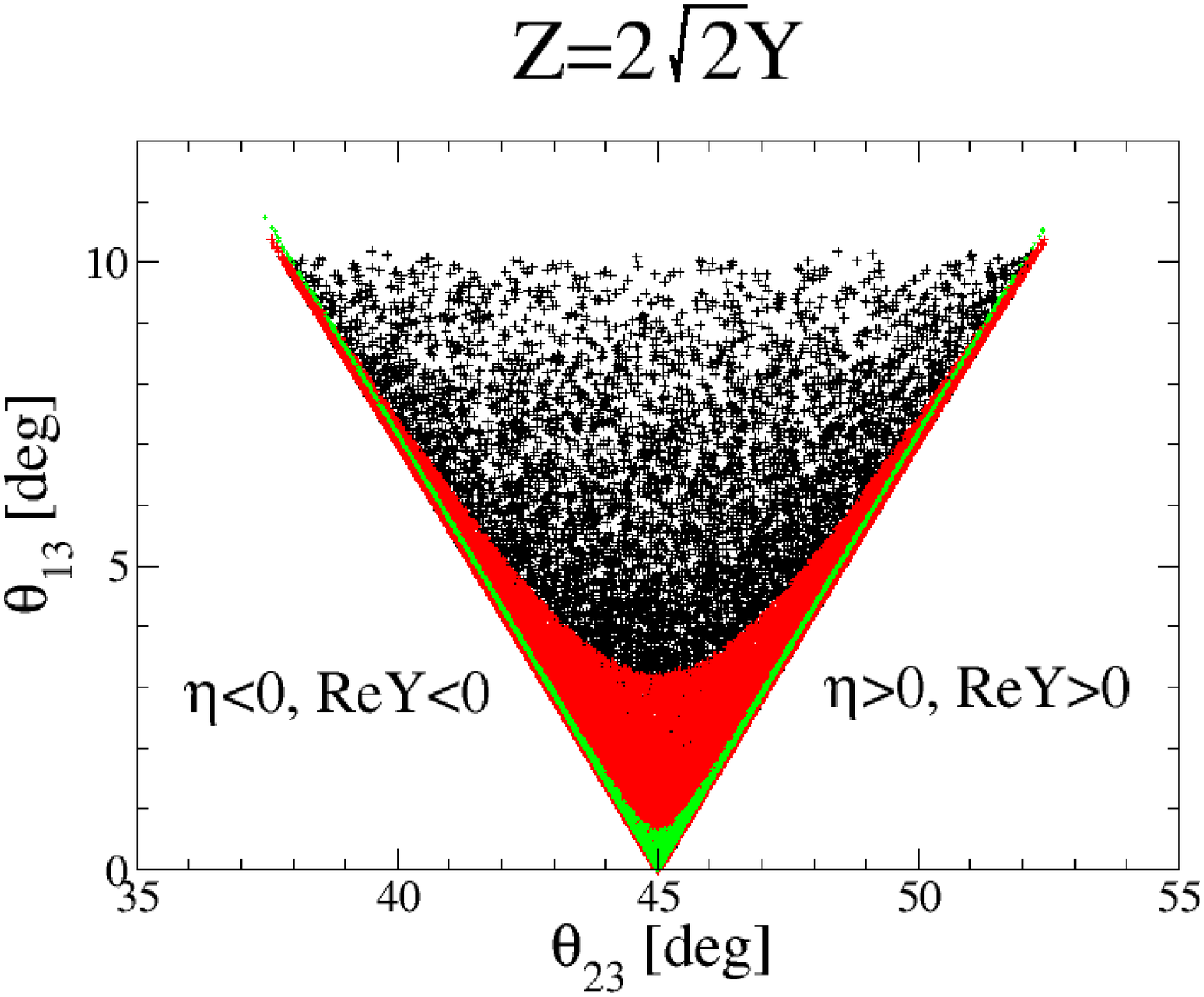}
\end{center}
\caption{\footnotesize 
Legend is the same as Fig. \ref{fig:atm-rct-efm-II-NH}, but for the inverted ordering case.
} \label{fig:atm-rct-efm-II-IH}
\end{figure}
In Figs. \ref{fig:atm-rct-efm-II-NH} and \ref{fig:atm-rct-efm-II-IH}, scatter plots of the $\theta_{13}^{} - \theta_{23}^{}$ plane are shown for the normal and inverted ordering cases.
The black, red, the green dots correspond to the cases of $m_{1(3)}^{}=0$, $0.03$, and $0.1~\ev$, respectively, in the case of normal (inverted) ordering.
It can be seen that the EFM relation becomes clear as the neutrino mass scale increases.
\item Case A-III: $Y=Y^*$, $Z=Z^*$ and $m_1^{}, m_2^{} \ll m_3^{}$.\\
Supposing $m_1^{}$ and $m_2^{}$ are negligibly small in comparison with $m_3^{}$, the dependences of Eq. (\ref{eq:13-23}) on the Majorana phases $\alpha$ and $\beta$ would be obscured.
In this case, Eqs. (\ref{eq:efm-I}), (\ref{eq:dMn-efm-I-Z0}), and  (\ref{eq:dMn-efm-I-ZY}) are again valid for real $Y$ and $Z$ without assuming specific values for  $\alpha$ and $\beta$.

\begin{figure}[t]
\begin{center}
\includegraphics[width=7.3cm]{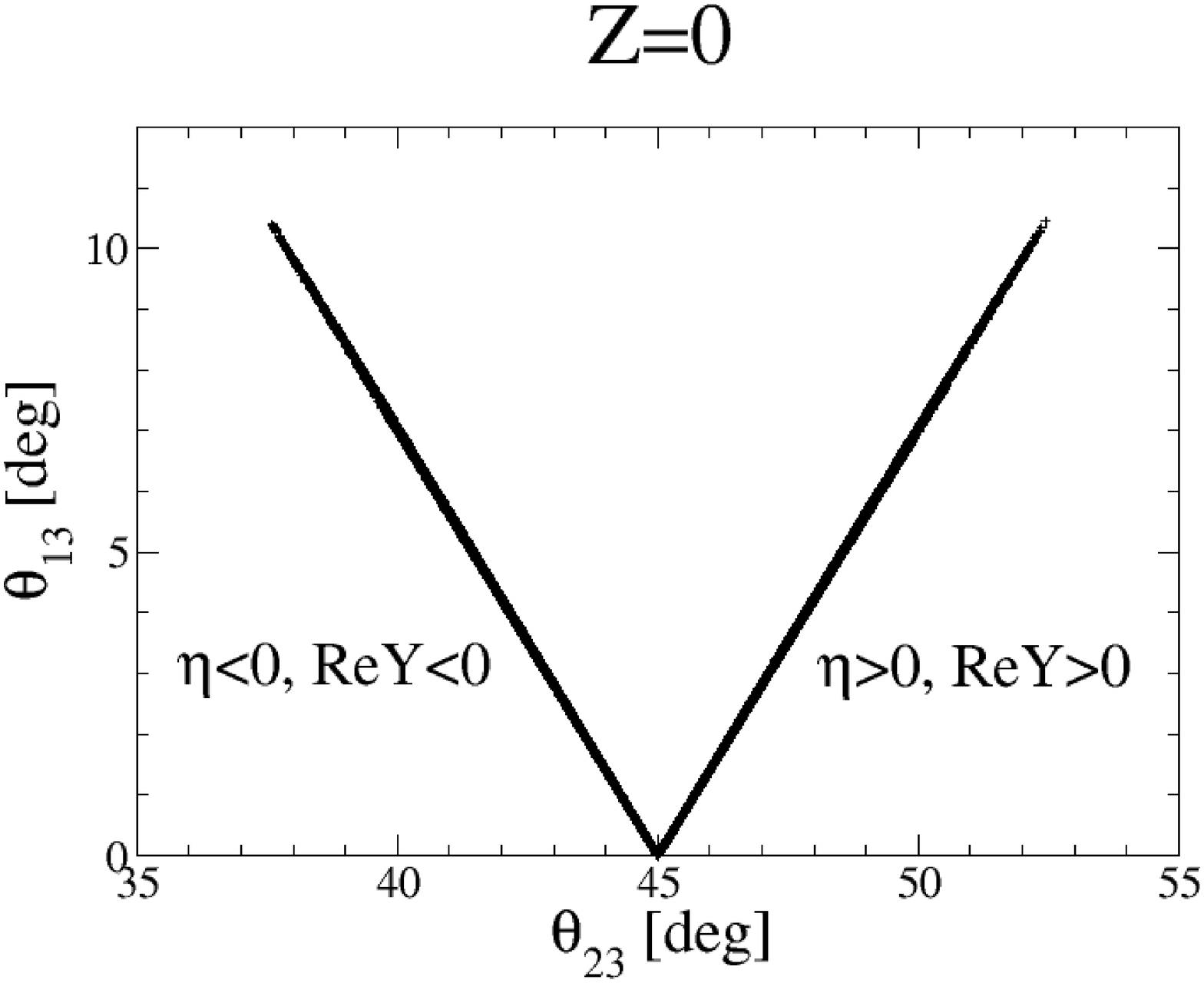}
\hspace{1.0cm}
\includegraphics[width=7.3cm]{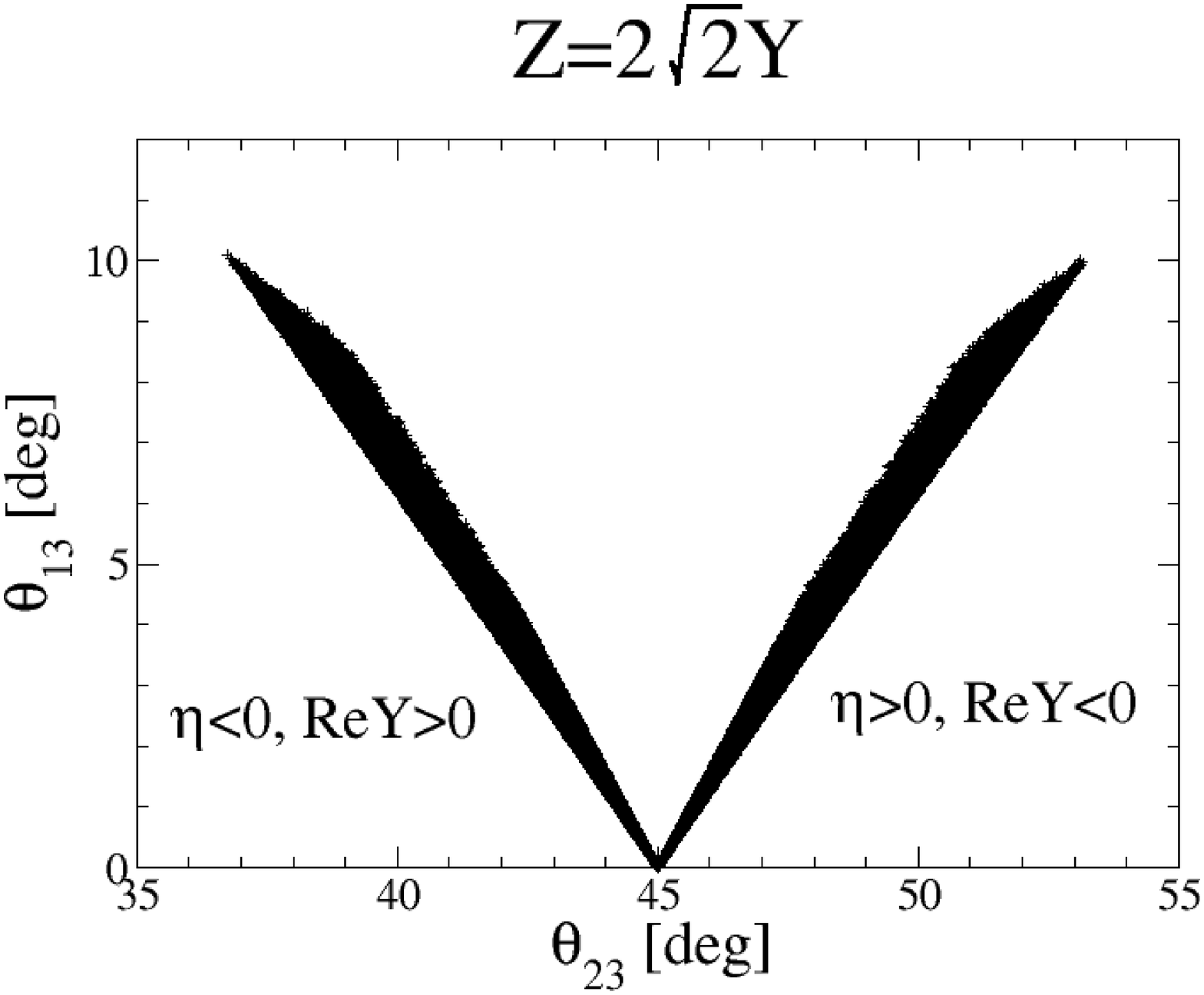} 
\end{center}
\caption{\footnotesize 
Legend is the same as Fig. \ref{fig:atm-rct-efm-I}, but for Case A-III.
} \label{fig:atm-rct-efm-III}
\end{figure}
In Fig. \ref{fig:atm-rct-efm-III}, $\theta_{13}$ and $\theta_{23}$ are numerically computed for the normal ordering case while fixing $m_1^{}$ at zero.
In the case of $Z=2\sqrt{2}Y$, the EFM relation is hazy as in Fig. \ref{fig:atm-rct-efm-I}, but this is more significant in the present case since we are looking at neutrino mass regions where errors of the approximation $\pm m_1^{} + m_3^{} = \pm m_2^{} + m_3^{}$ are maximized.

\end{itemize}

\subsection{$\eta=0$}
\begin{itemize}

\item Case B-I: $\alpha=\beta$ and $m_1^{} \simeq m_2^{} \gtrless m_3^{}$.\\
In the case where $m_1^{} \simeq m_2^{}$ and $\alpha=\beta$, the right-hand side of Eq. (\ref{eq:13-23}) turns out to be $\re[m_1^{} e^{i\alpha}(Y^*-\sqrt{2}Z^*) + m_3^{}(Y-\sqrt{2}Z)]$, which results in
\begin{eqnarray}
Z=\frac{Y}{\sqrt{2}}
\label{eq:max-I}
\end{eqnarray}
for $\eta=0$.

\begin{figure}[t]
\begin{center}
\includegraphics[width=7.3cm]{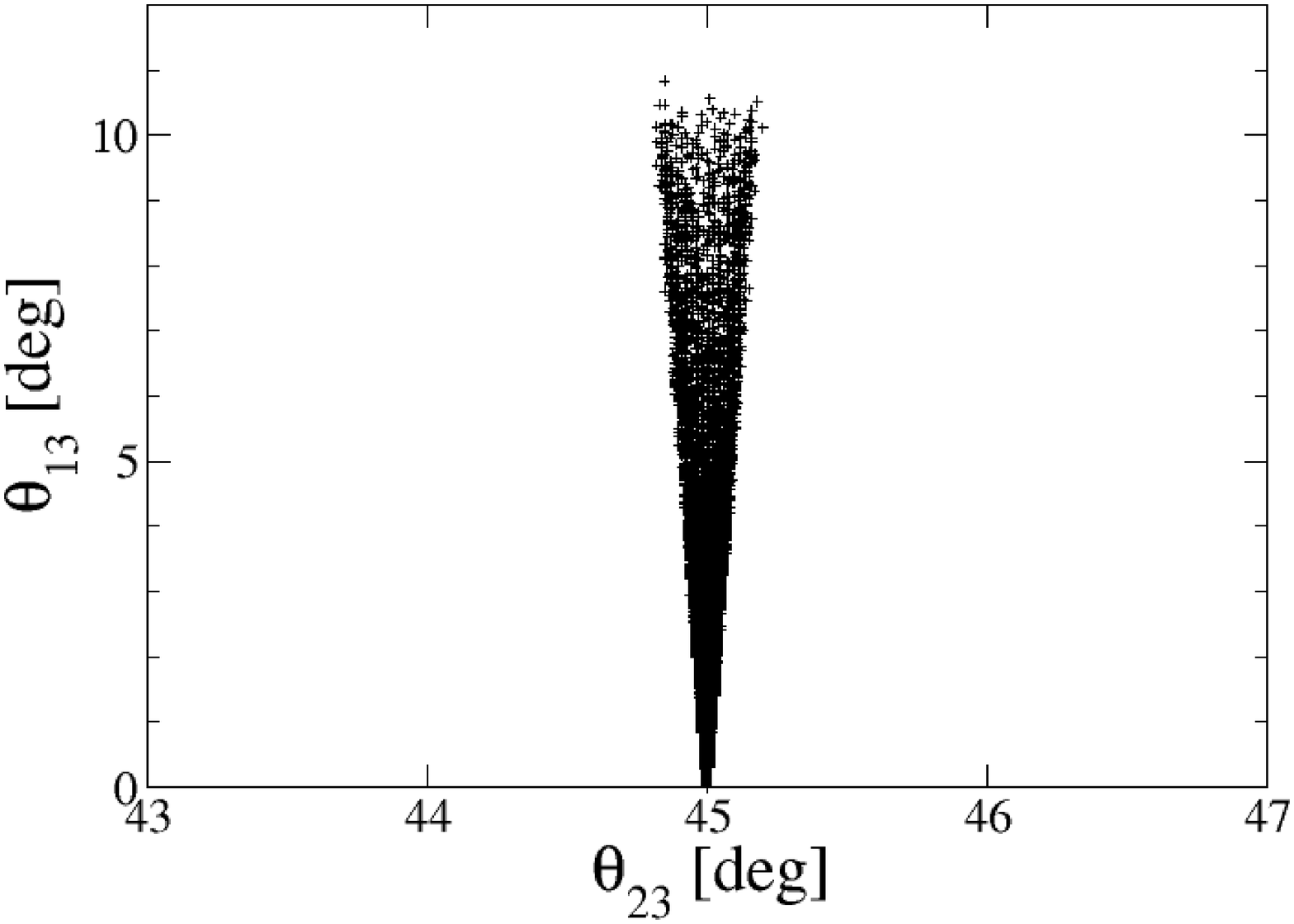}
\hspace{1.0cm}
\includegraphics[width=7.3cm]{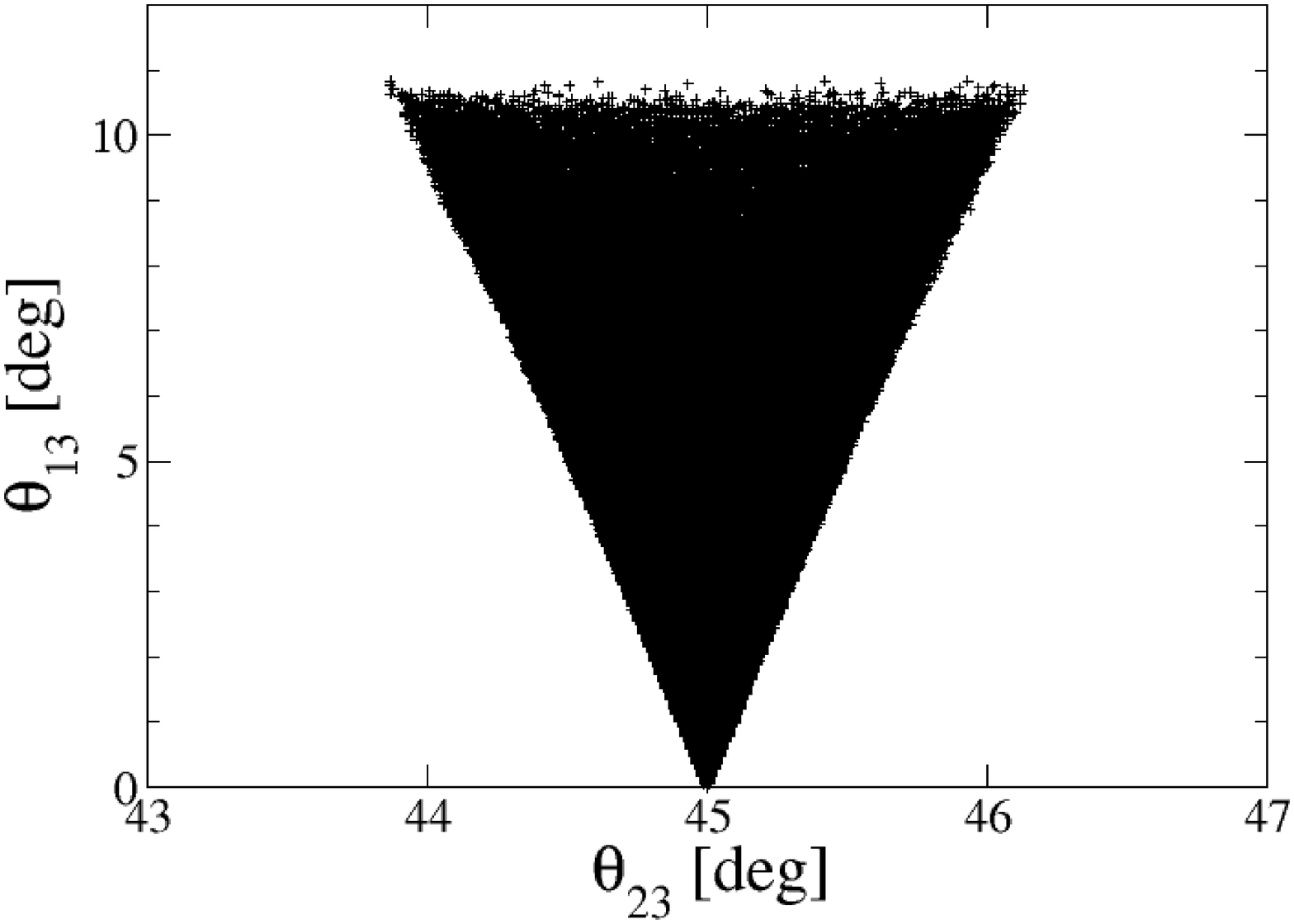} 
\end{center}
\caption{\footnotesize 
$\theta_{13}^{}$ as a function of $\theta_{23}^{}$ in the case of normal mass ordering for Case B-I.
$\Delta m_{12}^2$, $\Delta m_{23}^2$, and $\theta_{12}^{}$ are restricted to be within the 1$\sigma$ bounds.
In the left panel, $m_1^0$ is varied within $0.05 \sim 0.1~\ev$, while it is fixed at zero in the right panel.
} \label{fig:atm-rct-max-I}
\end{figure}
The results of numerical calculations for the case of normal mass ordering are displayed in Fig. \ref{fig:atm-rct-max-I}.
In the left panel, $m_1^{}$ is varied within $0.05 \sim 0.1~\ev$ in which $m_1^{} = m_2^{}$ is expected to be reasonable, while $m_1^{}$ is fixed at zero in the right panel in order to see how the stability of $\theta_{23}^{}$ is affected by $m_1^{} < m_2^{}$.
It can be seen that $\theta_{23}^{}$ scarcely departs from $45^\circ$ in the left panel, whereas $\theta_{23}^{}$ can be $45^\circ \pm 1^\circ$ in the right panel.
As for the inverted ordering case, the assumption $m_1^{} = m_2^{}$ is always valid, yielding almost the same figure as that in the left panel of Fig. \ref{fig:atm-rct-max-I}.
Hence, we refrain from showing it.

We here stress that Eq. (\ref{eq:max-I}) is equivalent to the condition found in Ref. \cite{max23}.
It can be seen from Eq. (\ref{eq:xyz}) that $Z=Y/\sqrt{2}$ (with $s_{13}^0=1/\sqrt{3}$) results in $(\delta\bar{M}_\nu)_{22}=(\delta\bar{M}_\nu)_{33}$; this is what was observed in Ref. \cite{max23}.

\item Case B-II: $-Y=Y^*$, $-Z=Z^*$, $\alpha=0$ or $\pi$ and $\beta=0$ or $\pi$.\\
Supposing that $Y$ and $Z$ are pure imaginary and that $\alpha$ and $\beta$ are equal to $0$ or $\pi$, the right-hand side of Eq. (\ref{eq:13-23}) turns out pure imaginary, and thus Eq. (\ref{eq:13-23}) is satisfied for $\eta=0$.

\begin{figure}[t]
\begin{center}
\includegraphics[width=7.3cm]{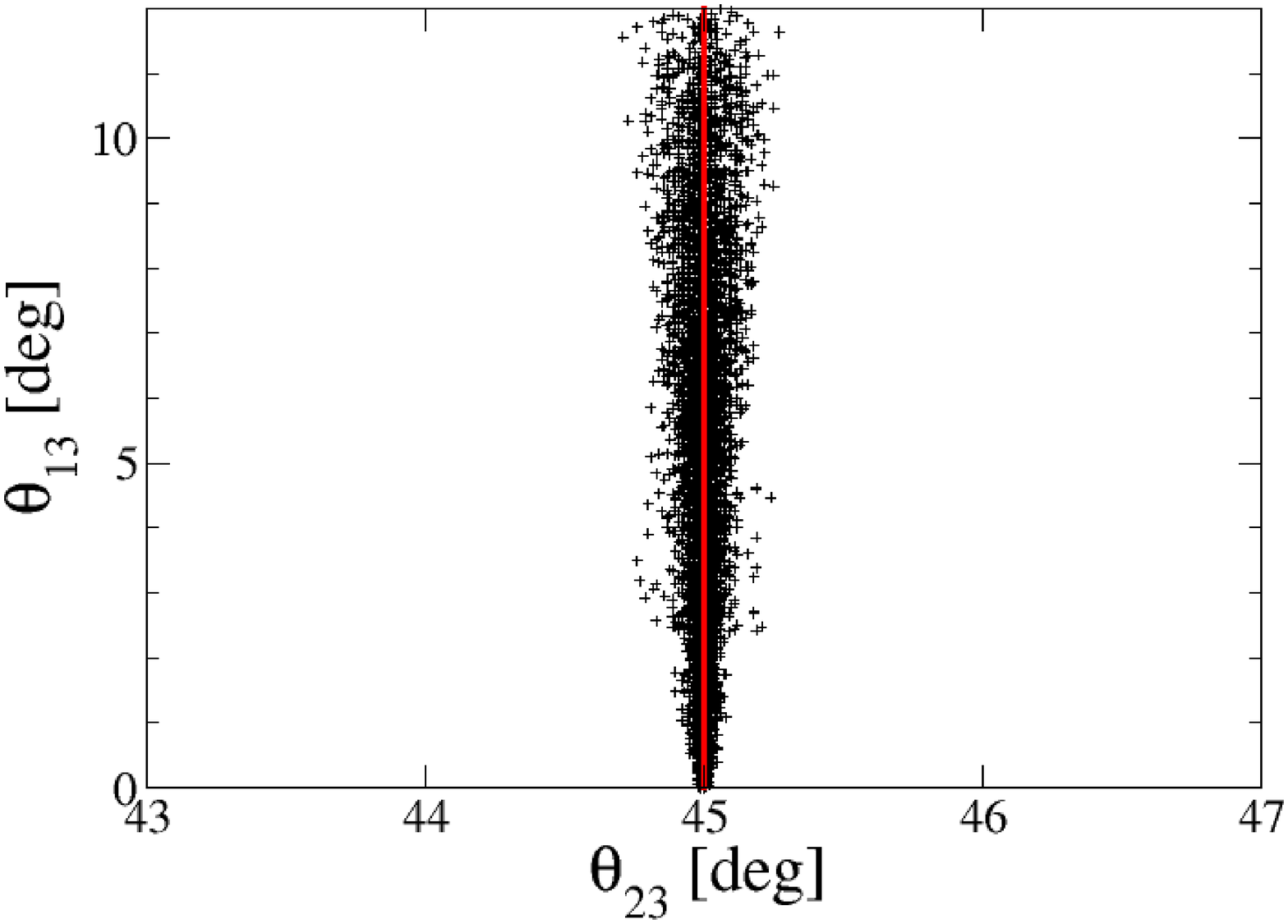}
\hspace{1.0cm}
\includegraphics[width=7.3cm]{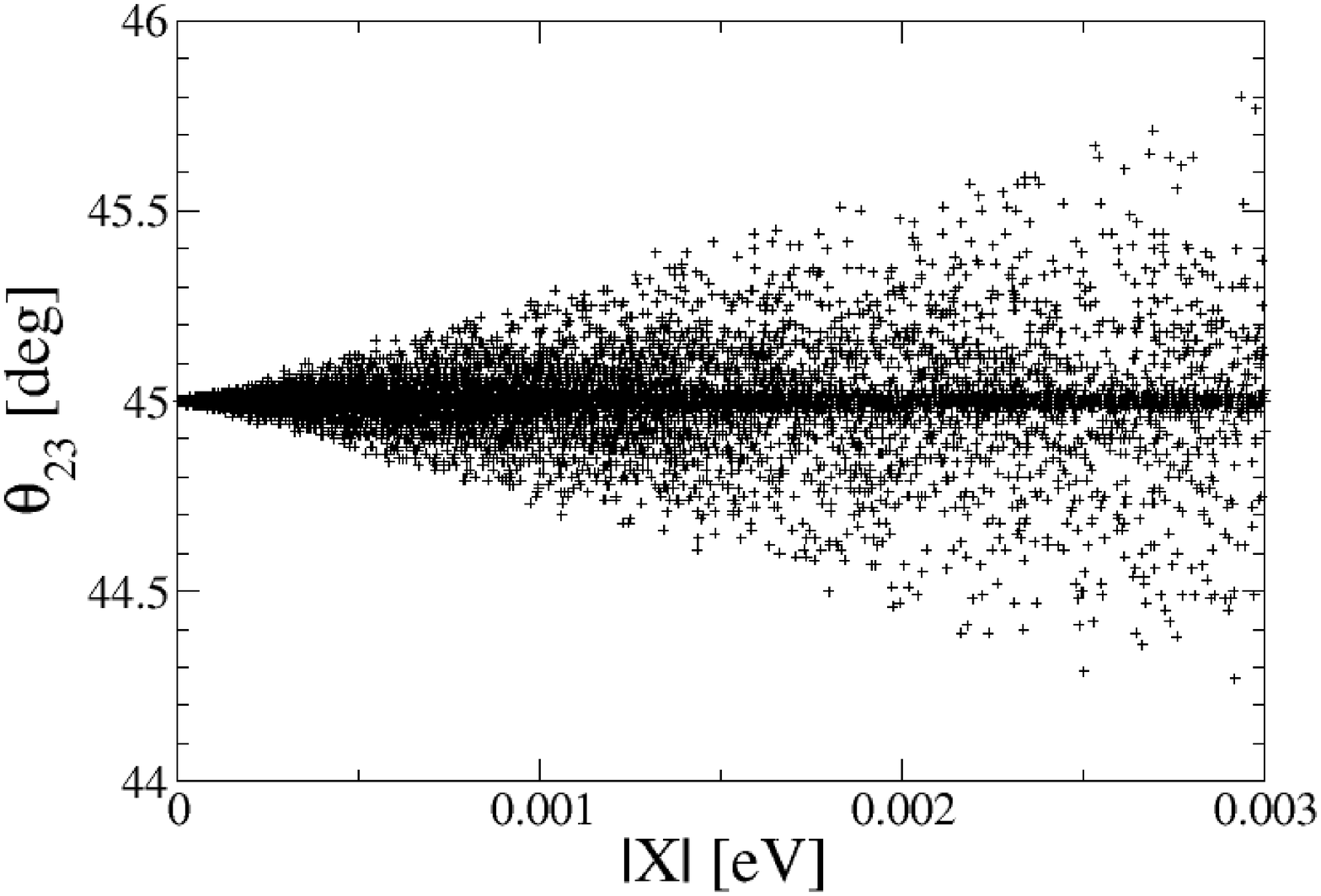} 
\end{center}
\caption{\footnotesize 
$\theta_{13}^{}$ as a function of $\theta_{23}^{}$ (left panel) and $\theta_{23}^{}$ as a function of $|X|$ (right panel) in the case of normal mass ordering for Case B-II.
$\Delta m_{12}^2$, $\Delta m_{23}^2$, and $\theta_{12}^{}$ are restricted to be within the 1$\sigma$ bounds.
In the left panel, $|X|$ is varied within $0 \sim 0.001~\ev$, and the red line corresponds to the case of $X=0$.
} \label{fig:atm-rct-max-II}
\end{figure}
In Fig. \ref{fig:atm-rct-max-II}, we plot $\theta_{13}^{}$ as a function of $\theta_{23}^{}$ (left panel) in the case of normal mass ordering.
This case prefers a somewhat small $X$.
In the right panel, we show the dependence of $\theta_{23}^{}$ on $|X|$; one can see that the deviations of $\theta_{23}^{}$ from $45^\circ$ can be large as $|X|$ increases.
The red vertical line in the left panel corresponds to the case of $X=0$.
The figures for the inverted ordering case are almost the same as Fig. \ref{fig:atm-rct-max-II}.

\end{itemize}

\section{Possible realizations of $\delta \bar{M}_\nu^{}$}
We here describe two possible ways of realizing Eqs. (\ref{eq:dMn-efm-I-Z0}) and (\ref{eq:max-I}).
The first one utilizes flavor symmetries, while the second one is based on radiative corrections.

As for Case B-II, it may be interesting to invoke the timeon mechanism \cite{timeon}, which is a mechanism of spontaneous CP violation triggered by the vacuum expectation value (VEV) of a pseudo-scalar.
Eq. (\ref{eq:dMn-efm-I-ZY}) could be explained by flavor symmetries with peculiar VEV alignments.
We do not discuss these in the present study.

\subsection{Flavor symmetry approach}
It is well known \cite{resdS4} that the TBM mixing can be derived from a neutrino mass matrix invariant under the following transformations:
\begin{eqnarray}
G_1^{}=
\bmx{ccc}
1 & 0 & 0 \\
0 & 0 & 1 \\
0 & 1 & 0
\emx,~~~
G_2^{}=\frac{1}{3}
\bmx{rrr}
 1 &-2 &-2 \\
-2 & 1 &-2 \\
-2 &-2 & 1
\emx, \nonumber
\end{eqnarray}
and that $G_1^{}$ and $G_2^{}$ belong to relatively small finite groups.
We show that $G_2^{}$ is still preserved if Eq. (\ref{eq:dMn-efm-I-Z0}), $Z=0$, holds.
Let us begin with the most general $G_2^{}$ invariant $\delta\bar{M}_\nu^{}$:
\begin{eqnarray}
\delta\bar{M}^{G_2}_\nu=
\bmx{ccc}
-b+c+d & b & a \\
 b & a-b+c & d \\
 a & d & c
\emx. \nonumber
\end{eqnarray}
Here, $c$ and $d$ can be absorbed into $\bar{M}_\nu^0$, so we will omit them hereafter.
Then, after some easy calculations, one can check
\begin{eqnarray}
\delta\bar{M}^{G_2}_\nu
=V^0_{}~\frac{1}{2}
\bmx{ccc}
 -(a+3b) & 0 & \sqrt{3}(a-b)/2 \\
 0 & 2a & 0 \\
 \sqrt{3}(a-b) & 0 & (a-b)
\emx (V^0_{})^T_{},
\label{eq:dMg2}
\end{eqnarray}
where $V^0_{}$ is the TBM mixing.
The diagonal entries contribute to the masses, but they can be ignored as long as $a$ and $b$ are sufficiently small in comparison with $\bar{M}_\nu^0$.
In this sense, Eq. (\ref{eq:dMg2}) is equivalent to Eq. (\ref{eq:xyz}) with $X=Z=0$\footnote{One can directly prove that Eq. (\ref{eq:xyz}) with $X=Z=0$ preserves $G_2^{}$ by following the method in Ref. \cite{resdS4}.}.
Since $X$ is mainly responsible for small deviations of $\theta_{12}^{}$, it may not be difficult to yield a non-zero $X$ from other sources.
To conclude, Eq. (\ref{eq:dMn-efm-I-Z0}) can be obtained if one can break $G_1^{}$ while maintaining $G_2^{}$.

Modifying the TBM mixing in the presence of Eq. (\ref{eq:dMg2}) was carefully studied in Ref. \cite{stw} without focusing on a correlation between $\theta_{13}^{}$ and $\theta_{23}^{}$.
They also proposed a simple model based on an $A_4^{}$ flavor symmetry, so we do not address model building here.
We would just like to emphasize that $A_4^{}$ may be suitable for realizing Eq. (\ref{eq:dMg2}).
In most $A_4^{}$ models, one needs to accidentally induce $G_1^{}$ as it is not the element of $A_4^{}$.
As a result, $G_1^{}$ is usually broken by higher-dimensional operators.

Note that it is also possible to reveal symmetries that guarantee Eq. (\ref{eq:dMn-efm-I-ZY}), $Z=2\sqrt{2}Y$, or Eq. (\ref{eq:max-I}), $Z=Y/\sqrt{2}$.
It is, however, unclear which group the symmetries belong to.
Therefore, we do not go into detail.

\subsection{Radiative corrections}
We consider a hybrid neutrino mass scheme composed of tree-level and one-loop operators.
The tree-level neutrino mass matrix is assumed to take the form of $\bar{M}_\nu^{0}$ in the diagonal basis of the charged lepton mass matrix, leading to TBM mixing at tree level.
For this setup, we insert two types of one-loop radiative corrections and show that they may be useful to realize Eqs. (\ref{eq:dMn-efm-I-Z0}) and (\ref{eq:max-I}).

The first example was proposed in Ref. \cite{fqc}, in which we have
\begin{eqnarray}
\delta \bar{M}_\nu^{} =
\frac{\bar{M}_\nu^0 D_\ell^2 + D_\ell^2 \bar{M}_\nu^0}{v_{\rm ew}^2} \times I^{\rm loop}_{}
\label{eq:fqc}
\end{eqnarray}
at the one-loop level, where $D_\ell^{}=\diag(m_e^{},m_\mu^{},m_\tau^{})$, $v_{\rm ew}^{}=174~\gev$, and $I^{\rm loop}_{}$ is a dimensionless parameter including some loop factors.
The details of $I^{\rm loop}_{}$ depend on the model details, so that we leave it arbitrary.
If $\alpha=0$ and $\beta=\pi$ in the case of the quasi-degenerate mass spectrum, Eq. (\ref{eq:fqc}) turns out to be
\begin{eqnarray}
\delta \bar{M}_\nu^{} 
&\simeq &
\frac{m_\nu^0}{3} \left[
m_\mu^2
\bmx{rrr}
0 & -2 & 0 \\
-2 & 2 & -2 \\
0 & -2 & 0
\emx
+m_\tau^2
\bmx{rrr}
0 & 0 & -2 \\
0 & 0 & -2 \\
-2 & -2 & 2
\emx
\right]\frac{I^{\rm loop}}{v_{ew}^2} \nonumber \\
&=&
V^0_{}
\left[
\frac{m_\nu^0}{3}
\bmx{ccc}
 m_\mu^2+m_\tau^2 & 0 & \sqrt{3}(m_\mu^2-m_\tau^2) \\
 0 & -2(m_\mu^2+m_\tau^2) & 0 \\
 \sqrt{3}(m_\mu^2-m_\tau^2) & 0 & 3(m_\mu^2+m_\tau^2)
\emx \frac{I^{\rm loop}}{v_{ew}^2} 
\right]
(V^0_{})^T_{}
\label{eq:dMn-fqc}
\end{eqnarray}
where $m_\nu^0 \equiv m_1^0 \simeq m_2^0 \simeq m_3^0$, and we have omitted the terms proportional to $m_e^2$.
The diagonal elements can be embedded into $\bar{M}_\nu^{0}$, and this example may be applicable to Case A-II\footnote{Recall that there is no constraint on $\beta$ in the case of $Z=0$.}.

The second example is the so-called Zee-model given in Ref. \cite{zee}, which induces
\begin{eqnarray}
\delta \bar{M}_\nu^{} 
&=&
\bmx{ccc}
 0 & f_{e\mu}^{}(m_\mu^2 - m_e^2) & 
 f_{e\tau}^{}(m_\tau^2 - m_e^2) \\
 f_{e\mu}^{}(m_\mu^2 - m_e^2) & 0 & 
 f_{\mu\tau}^{}(m_\tau^2 - m_\mu^2) \\
 f_{e\tau}^{}(m_\tau^2 - m_e^2) & 
 f_{\mu\tau}^{}(m_\tau^2 - m_\mu^2) & 0
\emx \frac{1}{\Gamma} 
\equiv
\bmx{ccc}
0 & a & b \\
a & 0 & c \\
b & c & 0
\emx \nonumber \\
&=&
V^0_{} \frac{1}{3}
\bmx{ccc}
 -2a-2b+c & (a+b-2c)/\sqrt{2} & \sqrt{3}(b-a) \\
 (a+b-2c)/\sqrt{2} & 2(a+b+c) & \sqrt{3/2}(b-a) \\
 \sqrt{3}(b-a) & \sqrt{3/2}(b-a) & -3c
\emx
(V^0_{})^T_{}
\label{eq:dMn-zee}
\end{eqnarray}
at the one-loop level.
$f_{\alpha\beta}^{}$ are Yukawa couplings, and $\Gamma$ contains loop factors and possesses mass-dimension one.
It can be seen that Eq. (\ref{eq:max-I}), $Z=Y/\sqrt{2}$, is achieved.
Note that Eq. (\ref{eq:dMn-zee}) contains a non-zero $X$ as well.
Thus, one can simultaneously obtain corrections to $\theta_{12}^{}$.

\section{Summary}
TBM mixing may still be useful as a leading order one in the presence of small corrections, although exact TBM mixing was ruled out by the discovery of a non-zero $\theta_{13}^{}$.
Once this direction is adopted, one is immediately faced with the difficulties of revealing the structures and origins of the corrections.
In this work, employing $\theta_{23}^{} \simeq 45^\circ + \eta\theta_{13}^{}$ as a guide, we seek preferable correction terms to the charged lepton and neutrino mass matrices and discuss their origins.
The latest global fit of the neutrino oscillation parameters points to $\eta=\pm 1/\sqrt{2}$, whereas the results of the  $\nu_\mu^{}$ disappearance mode seem to favor $\eta=0$.
We have succeeded in deriving the general condition Eq. (\ref{eq:13-23}) for ensuring $\theta_{23}^{} \simeq 45^\circ + \eta\theta_{13}^{}$, but it has also been found that the condition is complicated by the individual neutrino mass and CP violating phases.
Since these observables have not been measured yet, we investigate the condition for a specific neutrino mass spectrum, e.g. the quasi-degenerate spectrum, and/or characteristic CP violating phases, e.g. vanishing Majorana phases.
Under such simplified environments, we arrive at Eqs. (\ref{eq:dMn-efm-I-Z0}) and (\ref{eq:dMn-efm-I-ZY}) for the case of $\eta=\pm 1/\sqrt{2}$, and Eq. (\ref{eq:max-I}) for $\eta=0$.
Furthermore, we discuss possible realizations of them based on flavor symmetries and one-loop radiative corrections.

\end{document}